\documentclass[aps,pra,preprint,showpacs]{revtex4}

\usepackage{graphicx}% Include figure files
\usepackage{bm}% bold math

\def\lesssim{\ \raise.3ex\hbox{$<$}\kern-0.8em\lower.7ex\hbox{$\sim$}\ }
\def\gesim{\ \raise.3ex\hbox{$>$}\kern-0.8em\lower.7ex\hbox{$\sim$}\ }

\begin{document}
%\preprint{}

%Title of paper
\title{BCS-BEC crossover and effects of density fluctuations in a two-component Fermi gas loaded on an optical lattice}
\author{H. Tamaki$^1$, Y. Ohashi$^{2,3}$, K. Miyake$^{1}$}
\affiliation{
$^1$Department of Materials Engineering Science, Graduate School of Engineering Science, Osaka University, Toyonaka, Osaka 560-8531, Japan\\
$^2$Faculty of Science and Technology, Keio University, Hiyoshi, Yokohama 223-8522, Japan\\
$^3$CREST(JST), 4-1-8 Honcho, Saitama 332-0012, Japan}
\date{\today}

\begin{abstract}
We investigate the superfluid phase transition in a gas of Fermi atoms
 loaded on a three-dimensional optical lattice. When the lattice
 potential is strong, this system can be well described by an attractive
 Hubbard model. In this model, we calculate the superfluid phase
 transition temperature $T_{\rm c}$, including both superfluid and (spin
 and charge) density fluctuations within the self-consistent $t$-matrix
 theory and fluctuation exchange approximation, respectively. Since we
 treat these fluctuations in a consistent manner, our theory satisfies
 the required particle-hole symmetry over the entire BCS-BEC crossover
 region. We show that charge density fluctuations compete against
 superfluid fluctuations near the half-filling, leading to the
 suppression of $T_{\rm c}$. As a result, the maximum $T_{\rm c}$ is
 obtained away the half-filling. Since the strong density fluctuations
 originate from the nesting property of the Fermi surface at the half
 filling (which is absent in a uniform gas with no lattice potential),
 our results would be useful in considering lattice effects on
 strong-coupling superfluidity.
\end{abstract}


\maketitle

%%%%%%%%%%%%%%%%%%%%%%%%%%%%%%%%%%%%%%%%%%%%%%%%%%%%%%%%%%%%%%%%%%%%%%%%%%%%%%%
\section{introduction}
%%%%%%%%%%%%%%%%%%%%%%%%%%%%%%%%%%%%%%%%%%%%%%%%%%%%%%%%%%%%%%%%%%%%%%%%%%%%%%%
\par
Optical lattice is an artificial lattice produced by standing wave of
laser light\cite{Pitaevskii}. In a cold atom gas loaded on an optical
lattice, atoms feel a periodic potential due to the Stark effect. When
the lattice potential is strong, this system can be well described by
the Hubbard model, where atoms are hopping between nearest-neighbor
sites, interacting with each other when they meet at the same lattice
site. Since the Hubbard model is a fundamental model in condensed matter
physics, it is expected that various topics discussed in this field may
be solved by using the optical lattice system. Indeed, the
superfluid-Mott insulator transition has been observed in a $^{87}$Rb
lattice Bose gas\cite{Greiner,Esslinger}. More recently, the superfluid
state has been also realized in a $^6$Li lattice Fermi
gas\cite{chin06,Miller}. 
\par
Besides the optical lattice, a tunable interaction associated with a
Feshbach resonance is also an advantage of cold atom
gases\cite{Timmermans,Holland}. Using this unique property, several
experimental groups\cite{Jin,Grimm,Ketterle,Thomas} have succeeded in
realizing Fermi superfluids and the BCS-BEC (Bose-Einstein condensation)
crossover\cite{Leggett,Nozieres,Tokumitu,Melo,Haussmann94,Haussmann,Ohashi,Levin}
in the absence of optical lattice. The BCS-BEC crossover is a very
interesting phenomenon, because one can study the weak-coupling
BCS-state and the BEC of tightly bound molecules in a unified manner by
varying the strength of a pairing interaction. Since this tunable
pairing interaction also works in an optical lattice, it is interesting
to examine how the BCS-BEC crossover phenomenon is observed in the
optical lattice system. We briefly note that, in the Hubbard model,
interaction effects are parametrized by the scaled interaction $U/t$
(where $U$ and $t$ represent an on-site pairing interaction and
nearest-neighbor hopping, respectively). Thus, in addition to the direct
tuning of the pairing interaction by a Feshbach resonance, continuous
change from the weak- to strong-coupling regime can be also realized by
adjusting the hopping parameter $t$ by tuning the intensity of laser
light producing the optical lattice.
\par
The BCS-BEC crossover in the attractive Hubbard model has been discussed
in superconductivity literature\cite{Micnus}, in connection to
strongly-correlated electron systems. Nozi\`eres and
Schmitt-Rink\cite{Nozieres} pointed out the mass enhancement of tightly
bound molecules in the strong-coupling BEC regime, because of virtual
dissociation of a bound molecule during hopping between lattice
sites. This mass enhancement is expected to decrease the superfluid
phase transition temperature $T_{\rm c}$ in the BEC regime, which has
been theoretically confirmed by the self-consistent $t$-matrix
theory\cite{keller99}, dynamical mean-field theory\cite{keller01}, and
quantum Monte-Carlo simulation\cite{sewer02}. Competition between
pairing fluctuations and charge density wave (CDW)
fluctuations\cite{note} near the half-filling (which comes from the
nesting property of the square-shape Fermi surface at the half-filling)
has been also studied\cite{taraph95}. In two-dimension,
Refs. \cite{Scalettar,deisz02} pointed out that this competition leads
to vanishing $T_{\rm c}$ at the half-filling. For more details, we refer
to Ref. \cite{Micnus}.
\par
In this paper, we investigate the superfluid phase transition in a gas
of Fermi atoms loaded on a three-dimensional cubic optical
lattice. Treating this system as the Hubbard model, we calculate $T_{\rm
c}$ in the BCS-BEC crossover region, including pairing fluctuations, as
well as CDW and spin density wave (SDW) fluctuations in a consistent
manner within the self-consistent $t$-matrix approximation
(SCTA)\cite{Haussmann94,Haussmann,engel02} and fluctuation exchange
approximation (FLEX)\cite{deisz02}, respectively. Our theory satisfies
the required particle-hole symmetry over the entire BCS-BEC crossover
region. While a finite $T_{\rm c}$ is obtained at the half-filling in
contrast to the two-dimensional case\cite{Scalettar}, the superfluid
phase transition is shown to be strongly influenced by CDW fluctuations
near the half-filling. The resulting $T_{\rm c}$ takes the maximum
value, not at the half-filling, but around the quarter filling. Since
strong CDW fluctuations are characteristic of the lattice system we
consider in this paper, the observation of the filling dependence of
$T_{\rm c}$ would be an interesting problem. 

\par
This paper is organized as follows. In Sec. II, we explain our
formulation. The self-consistent $t$-matrix approximation (SCTA) for
pairing fluctuations and the fluctuation exchange approximation (FLEX)
for CDW and SDW fluctuations are explained. In Sec. III, we present our
numerical results for the superfluid phase transition temperature, only
taking into account pairing fluctuations. We examine effects of CDW and
SDW fluctuations on the superfluid phase transition in
Sec. IV. Throughout this paper, we set $\hbar=k_B=1$.

%%%%%%%%%%%%%%%%%%%%%%%%%%%%%%%%%%%%%%%%%%%%%%%%%%%%%%%%%%%%%%%%%%%%%%%%%%%%%%%
\section{Formulation}
%%%%%%%%%%%%%%%%%%%%%%%%%%%%%%%%%%%%%%%%%%%%%%%%%%%%%%%%%%%%%%%%%%%%%%%%%%%%%%%
\par
We consider a two-component Fermi gas in a three-dimensional cubic
optical lattice. In superfluid Fermi gases, all the current experiments
are using a broad Feshbach resonance\cite{Jin,Ketterle,Grimm,Thomas}. In
this case, details of the Feshbach resonance is known to be not
important as far as we consider the interesting BCS-BEC crossover
region, so that we can safely consider this system using the ordinary
BCS model. In addition, as mentioned in the introduction, a Fermi gas in
an optical lattice can be well described by the Hubbard model when the
lattice potential is strong. Under these conditions, we consider the
attractive Hubbard model described by the Hamiltonian
\begin{eqnarray}
H=-t\sum_{(i,j),\sigma}(c^{\dagger}_{i,\sigma} c_{j,\sigma}
 + \mathrm{h.c.})-U\sum_{i}n_{i\uparrow}n_{i\downarrow}
 - \mu\sum_{i,\sigma} n_{i,\sigma}.
 \label{eq.1}
\end{eqnarray}
Here, $c_{j,\sigma}$ is the annihilation operator of a Fermi atom at the
$j$-th lattice site, where the pseudo-spin $\sigma=\uparrow,\downarrow$
describes two atomic hyperfine states. $t$ is the hopping matrix element
between nearest-neighbor sites, and the summation $(i,j)$ in the first
term is taken over nearest-neighbor
pairs. $n_{i\sigma}=c^\dagger_{i\sigma}c_{i\sigma}$ is the number
operator. The on-site pairing interaction $-U~(<0)$ is implicitly
assumed to be tunable by using a Feshbach resonance. $\mu$ is the Fermi
chemical potential. In Eq. (\ref{eq.1}), we have neglected effects of a
harmonic trap, for simplicity. 
\par
%%%%%%%%%%%%%%%%%%%%%%%%%%%%%%%%%%%%%%%%%%%%%%%%%%%%%%%%%%%%%%%%%%%%%%%%%%%%%%%
\begin{figure}
\includegraphics[scale=1.3]{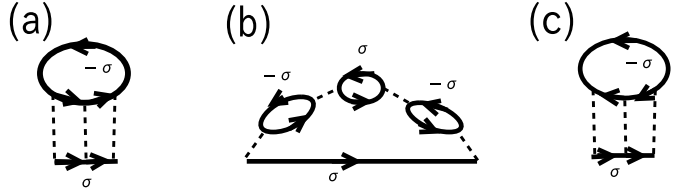}
\caption{
Self-energy corrections describing (a) pairing fluctuations, (b) CDW fluctuations and SDW fluctuations of the $z$-component, and (c) SDW fluctuations of the $x$- and $y$-component. The solid line and dashed line describe the single-particle Green's function $G$ and the attractive interaction $-U$, respectively.
\label{fig1}
}
\end{figure}
%%%%%%%%%%%%%%%%%%%%%%%%%%%%%%%%%%%%%%%%%%%%%%%%%%%%%%%%%%%%%%%%%%%%%%%%%%%%%%%
\begin{figure}
\includegraphics[scale=0.6]{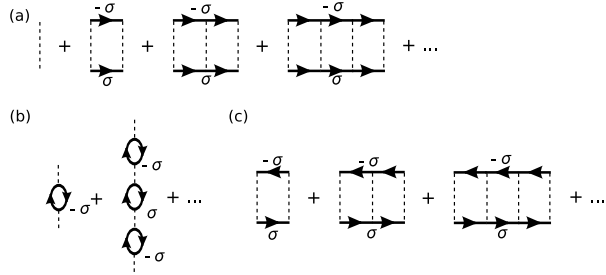}
\caption{
Vertex functions of (a) particle-particle channel $\Gamma_{\rm pp}$ and (b), (c) particle-hole channel $\Gamma_{\rm ph}^{\rm s,d}$. In our calculation, the lowest order term in terms of $U$ is included in $\Gamma_{\rm pp}$.
 \label{fig2}
}
\end{figure}
%%%%%%%%%%%%%%%%%%%%%%%%%%%%%%%%%%%%%%%%%%%%%%%%%%%%%%%%%%%%%%%%%%%%%%%%%%%%%%%
\par
We calculate the superfluid phase transition temperature $T_{\rm c}$,
extending SCTA developed by Haussmann\cite{Haussmann94,Haussmann} so as to
include CDW and SDW fluctuations within FLEX. Effects of these
fluctuations are described by the self-energy $\Sigma({\bm
k},i\omega_m)$ in the single-particle Green's function $G({\bm
k},i\omega_m)$,
\begin{equation}
G(\bm{k},i\omega_m)=
{1 \over
i\omega_m-\varepsilon_{\bf k}+\mu-\Sigma({\bf k},i\omega_m)},
\label{eq2}
\end{equation}
where $\varepsilon_{\bf k}\equiv-2t[\cos k_x+\cos k_y+\cos k_z]-\mu$ is
the kinetic energy of a Fermi atom measured from the chemical potential
$\mu$ (where the lattice constant is taken to be unity). $\omega_m$ is
the fermion Matsubara frequency. 
\par
Figure \ref{fig1}(a) shows the self-energy correction coming from
pairing fluctuations ($\equiv \Sigma_{\rm pp}$). Summing up this type of
diagrams within SCTA, we obtain
\begin{equation}
\Sigma_{\rm pp}({\bm k},i\omega_m)=
{1 \over \beta}\sum_{{\bm q},i\nu_n}
\Gamma_{\rm pp}({\bm q},i\nu_n)G({\bm q}-{\bm k},i\nu_n-i\omega_m),
\label{eq3}
\end{equation}
where $\nu_n$ is the boson Matsubara frequency.
Hereafter, the factor $1/N$ of lattice sites number in front of momentum
summation is abbreviated for simple presentation.
$\Gamma_{\rm pp}$ is the particle-particle scattering vertex diagrammatically described by Fig.\ref{fig2}(a). The result is
\begin{equation}
\Gamma_{\rm pp}({\bm q},i\nu_n)=
-{U \over 1-U\Pi_{\rm pp}({\bm q},i\nu_n)},
\label{eq4}
\end{equation}
where,
\begin{equation}
\Pi_{\rm pp}({\bm q},i\nu_n)=
{1 \over \beta}\sum_{{\bf k},i\omega_m}
G({\bm k},i\omega_m)G({\bf q}-{\bm k},i\nu_n-i\omega_m)
\label{eq5}
\end{equation}
is a correlation function describing fluctuations in the Cooper channel.
\par
In Fig.\ref{fig1}, panels (b) and (c) describe fluctuations in the
particle-hole channel within FLEX. Panel (b) involves both CDW
fluctuations and longitudinal SDW fluctuations. Panel (c) involves
transverse SDW fluctuations. Summing up these diagrams, we obtain the
self-energy corrections associated with CDW fluctuations $(\equiv
\Sigma_{\rm ph}^{\rm d})$ and SDW fluctuations ($\equiv \Sigma_{\rm
ph}^{\rm s})$ as
\begin{eqnarray}
\Sigma_{\rm ph}^{\rm d,s}({\bm k},i\omega_m)
&=&
-{U \over \beta}\sum_{{\bf q},i\nu_n}
\Pi_{\rm ph}({\bm q},i\nu_n)\Gamma_{\rm ph}^{\rm d,s}({\bm
q},i\nu_n)G({\bm k}-{\bm q},i\omega_m-i\nu_n)\nonumber\\
&\equiv&
{1 \over \beta}\sum_{{\bf q},i\nu_n}
V_{\rm ph}^{\rm d,s}({\bm q},i\nu_n)G({\bm k}-{\bm q},i\omega_m-i\nu_n),
\label{eq6}
\end{eqnarray}
where $V_{\rm ph}^{\rm d,s}$ is introduced for saving the computational time by
performing the fast Fourier transformation (FFT), which is explained in Sec. III. 
Here, the vertex functions $\Gamma_{\rm ph}^{\rm d}$ and $\Gamma_{\rm
ph}^{\rm s}$ are obtained from the sum of the diagrams shown in
Figs.\ref{fig2}(a) and \ref{fig2}(b), and their expressions are given by
\begin{eqnarray}
\Gamma_{\rm ph}^{\rm d}(\bm{q},i\nu_n)=
-{1 \over 2}
{
U^2\Pi_{\rm ph}({\bm q},i\nu_n) 
\over 
1-U\Pi_{\rm ph}({\bm q},i\nu_n)
}.
\label{eq7}
\end{eqnarray}
\begin{eqnarray}
\Gamma_{\rm ph }^{\rm s }({\bm q},i\nu_n)=
{3 \over 2}
{U^2\Pi_{\rm ph}({\bm q},i\nu_n)
\over 
1+U\Pi_{\rm ph}({\bm q},i\nu_n)
}.
\label{eq8}
\end{eqnarray}
In Eqs. (\ref{eq7}) and (\ref{eq8}), the correlation function $\Pi_{\rm ph}({\bm k},i\nu_n)$ describes fluctuations in the particle-hole channel, having the form
\begin{equation}
\Pi_{\rm ph}({\bm q},i\nu_n)=-
{1 \over \beta}\sum_{{\bm k},i\omega_m}
G({\bm k},i\omega_m)G({\bf k}-{\bm q},i\omega_m-i\nu_n).
\label{eq9}
\end{equation}
The superfluid phase transition temperature $T_{\rm c}$ is determined from the Thouless criterion\cite{Haussmann94,Haussmann}, stating that the superfluid phase transition occurs when the particle-particle scattering vertex $\Gamma_{\rm pp}({\bm q},i\nu_n)$ has a pole at ${\bm q}=\nu_n=0$. Using this, we obtain the equation for $T_{\rm c}$ as
\begin{equation}
1=U\Pi_{\rm pp}({\bm q}=0,i\nu_n=0).
\label{eq10}
\end{equation}
We note that Eq. (\ref{eq10}) is affected by CDW and SDW fluctuations through the self-energy $\Sigma=\Sigma_{\rm pp}+\Sigma_{\rm ph}^{\rm d}+\Sigma_{\rm ph}^{\rm s}$ in the Green's function.
\par
In the weak-coupling BCS regime, we may set $\mu=\varepsilon_{\rm F}$ (where $\varepsilon_{\rm F}$ is the Fermi energy) in Eq. (\ref{eq10}). However, the chemical potential is known to deviate from $\varepsilon_{\rm F}$, as one approaches the strong-coupling BEC regime\cite{Leggett,Nozieres}. This strong-coupling effect is taken into account by considering the equation for the filling number $n$ (which gives the number of atoms per lattice site), given by
\begin{equation}
n={2 \over \beta}\sum_{{\bm k},i\omega_m}e^{i\omega_m\delta}G({\bm k},i\omega_m),
\label{eq11}
\end{equation}
where $\delta$ is an infinitesimal positive number. We solve the coupled equations (\ref{eq10}) and (\ref{eq11}) to determine $T_{\rm c}$ and $\mu$ self-consistently for a given $U$ and $n$.
\par
At the half-filling $n=1$, the superfluid state and CDW are degenerate in the sense that they have the same phase transition temperature\cite{Micnus,taraph95}. The CDW phase transition is characterized by the divergence of the charge susceptibility $\chi_{\rm CDW}({\bm Q})$ with the momentum ${\bm Q}=(\pi,\pi,\pi)$. In the random phase approximation, $\chi_{\rm CDW}$ is given by
\begin{eqnarray}
\chi_{\rm CDW}({\bm Q}=(\pi,\pi,\pi))=
\frac{1}{2}
\frac{\Pi_{\rm ph}({\bm Q},0)}{1-U\Pi_{\rm ph}({\bm Q},0)},
\label{eq12}
\end{eqnarray}
where the correlation function $\Pi_{\rm ph}({\bm Q},i\nu_n)$ is given by Eq. (\ref{eq9}). Comparing Eq. (\ref{eq12}) with Eq. (\ref{eq7}), we find that the CDW vertex function $\Gamma_{\rm ph}^{\rm d}({\bm Q},i\nu_n=0)$ also diverges at $T_{\rm c}$. This CDW instability at the superfluid phase transition temperature $T_{\rm c}$ is absent when $n\ne 1$ due to the absence of the perfect nesting of the Fermi surface. However, since the denominator $1-U\Pi_{\rm ph}({\bm Q},i\nu_n=0)$ in Eq. (\ref{eq12}) is still small at $T_{\rm c}$ near the half-filling, strong CDW fluctuations are expected when $n\sim 1$.
\par
In contrast, the SDW vertex function $\Gamma_{\rm ph}^{\rm s}$ in Eq. (\ref{eq8}) does not diverge at $T_{\rm c}$ even when $n=1$. Namely, spin fluctuations are weak in the attractive Hubbard model.
\par
%%%%%%%%%%%%%%%%%%%%%%%%%%%%%%%%%%%%%%%%%%%%%%%%%%%%%%%%%%%%%%%%%%%%%%%%%%%%%%%
\section{Effects of pairing fluctuations on $T_{\rm c}$ and $\mu$ in the BCS-BEC crossover region}
%%%%%%%%%%%%%%%%%%%%%%%%%%%%%%%%%%%%%%%%%%%%%%%%%%%%%%%%%%%%%%%%%%%%%%%%%%%%%%%\\par
In the following two sections, we show our numerical results obtained by solving the coupled equations (\ref{eq10}) and (\ref{eq11}). In this section, we first consider $T_{\rm c}$ in the BCS-BEC crossover, including pairing fluctuations only. Although one cannot actually ignore strong CDW fluctuations near the half-filling, examining this simple case is still useful in considering importance of CDW and SDW fluctuations. We separately discuss effects of CDW and SDW fluctuations in Sec. IV. 
\par
Before showing our results, we summarize the outline of computation. In solving the coupled equations (\ref{eq10}) and (\ref{eq11}), we use the fact that the self-energy $\Sigma=\Sigma_{\rm pp}+\Sigma_{\rm ph}^{\rm d}+\Sigma_{\rm ph}^{\rm s}$ in Eqs.(\ref{eq3}) and (\ref{eq6}), the correlation functions $\Pi_{\rm pp}$ and $\Pi_{\rm ph}$ in Eqs. (\ref{eq5}) and (\ref{eq9}), and the number equation in Eq. (\ref{eq11}) have simple expressions in real space, as
\begin{eqnarray}
\Sigma({\bm r},\tau)
&=&
\Sigma_{\rm pp}({\bm r},\tau)+
\Sigma_{\rm ph}^{\rm d}({\bm r},\tau)+
\Sigma_{\rm ph}^{\rm s}({\bm r},\tau)
\nonumber
\\
&=&
\Gamma_{\rm pp}({\bm r},\tau)G(-{\bm r},-\tau)+
\Bigl[
 V_{\rm ph}^{\rm d}({\bm r},\tau)+
 V_{\rm ph}^{\rm s}({\bm r},\tau)
\Bigr]G({\bm r},\tau),
\label{eq13}
\end{eqnarray}
\begin{eqnarray}
\Pi_{\rm pp}({\bm r},\tau)=G({\bm r},\tau)G({\bm r},\tau),
\label{eq14}
\end{eqnarray}
\begin{eqnarray}
\Pi_{\rm ph}({\bm r},\tau)=-G({\bm r},\tau)G(-{\bm r},-\tau),
\label{eq15}
\end{eqnarray}
\begin{equation}
n=2G(\bm{r}=0,\tau=-\delta). 
\label{eq16}
\end{equation}
Here, ${\bm r}$ is the spatial position of a lattice site and $\tau$ is the imaginary time. The Fourier transformation is defined by
\begin{eqnarray}
\left\{
\begin{array}{ll}
\displaystyle
G({\bm r},\tau)={1 \over \beta}\sum_{{\bf k},i\omega_m}G({\bm k},i\omega_m)
e^{i({\bm k}\cdot{\bm r}-\omega_m\tau)},\\
\displaystyle
G({\bm k},i\omega_m)=\sum_{\bf r}\int_0^\beta d\tau G({\bm k},\tau)
e^{-i({\bm k}\cdot{\bm r}-\omega_m\tau)}.\\
\end{array}
\right.
\label{eq17}
\end{eqnarray}
To use Eqs. (\ref{eq13})-(\ref{eq16}), we employ the FFT method\cite{note2}. We discretize the momentum
region $0\le k_x,k_y,k_z\le\pi$ into $16\times16\times16$ cells. For the
frequency summations, we introduce a finite cutoff frequency
$\omega_{\rm max}=\pi T(2n_{\rm max}+1)$ for fermions and $\nu_{\rm
max}=2\pi T n_{\rm max}$ for boson, with $n_{\rm max}=512$. The values
of these cutoffs are chosen so as to be much larger than the band width $2zt$
(where $z=6$ is the coordination number of the simple cubic lattice), as
well as the magnitude of the pairing interaction $U$. To avoid effects
of these cutoff frequencies, we use the method discussed in
\cite{deisz02}. We explain the outline of this method in the Appendix.
\par

%%%%%%%%%%%%%%%%%%%%%%%%%%%%%%%%%%%%%%%%%%%%%%%%%%%%%%%%%%%%%%%%%%%%%%%%%%%%%%%
\begin{figure}
\includegraphics[scale=0.7]{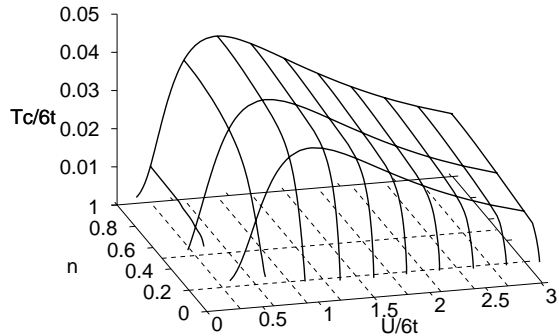}
\caption{Superfluid phase transition temperature $T_{\rm c}$ as a function of pairing interaction $U$ and filling number $n$. In this figure, and in Fig.\ref{fig4}, we only include pairing fluctuations described by $\Sigma_{\rm pp}$. Since $T_{\rm c}$ at the filling $2-n$ is the same as $T_{\rm c}$ at $n$ due to the particle-hole symmetry, we only show the result less than half-filling ($n\le 1$). 
\label{fig3}
}
\end{figure}
%%%%%%%%%%%%%%%%%%%%%%%%%%%%%%%%%%%%%%%%%%%%%%%%%%%%%%%%%%%%%%%%%%%%%%%%%%%%%%%

%%%%%%%%%%%%%%%%%%%%%%%%%%%%%%%%%%%%%%%%%%%%%%%%%%%%%%%%%%%%%%%%%%%%%%%%%%%%%%%
\begin{figure}
\includegraphics[scale=0.7]{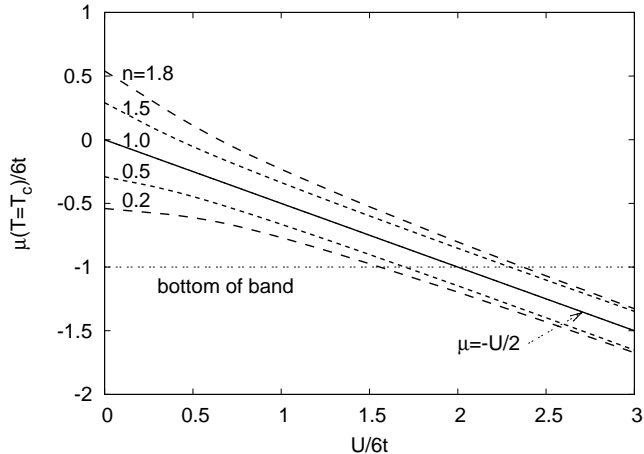}
\caption{Chemical potential $\mu$ as a function of $U$ at $T_{\rm c}$. $\mu=-U/2$ is the exact solution at the half-filling $n=1$. At $U=0$, apart from weak temperature effect, $\mu$ is almost equal to the Fermi energy $\varepsilon_{\rm F}$ for a given filling number $n$.
\label{fig4}
}
\end{figure}
%%%%%%%%%%%%%%%%%%%%%%%%%%%%%%%%%%%%%%%%%%%%%%%%%%%%%%%%%%%%%%%%%%%%%%%%%%%%%%%

\par
Figure \ref{fig3} shows the calculated $T_{\rm c}$ in the BCS-BEC crossover. In this calculation, we only include the self-energy $\Sigma_{\rm pp}$ associated with pairing fluctuations. Since the Hubbard model  has the particle-hole symmetry, the filling dependence of $T_{\rm c}$ is symmetric with respect to $n=1$ (although we do not show it explicitly). Namely, for a given $U$, Fig.\ref{fig3} shows that the maximum $T_{\rm c}$ is obtained at the half-filling $n=1$, while $T_{\rm c}$ vanishes at $n=2$ as in the case of $n=0$. 
\par
In the weak-coupling BCS regime ($U/6t\lesssim 1$), $T_{\rm c}$ is an increasing function of $U$. This behavior agrees with the well-known mean-field BCS result, 
\begin{equation}
T_{\rm c}\propto t e^{-{1 \over N(0)U}},
\label{eq18}
\end{equation}
where $N(0)$ is the density of states at the Fermi level. In this regime, since $T_{\rm c}$ is comparable to the binding energy $E_{\rm bind}$ of a Cooper pair at $T=0$, the increase of $T_{\rm c}$ reflects the enhancement of $E_{\rm bind}$ as one approaches the strong-coupling regime.
\par
For a given filling number $n$, the maximum $T_{\rm c}$ ($\equiv T_{\rm c}^{\rm max}$) is obtained around the intermediate coupling region $U/6t\sim 1$. At the half-filling ($n=1$), we obtain $T_{\rm c}^{\rm max}=0.042$ at $U/6t=0.9$. $T_{\rm c}$ then decreases as one further increases the magnitude of $U$. The decrease of $T_{\rm c}$ in the BEC regime is characteristic of the BCS-BEC crossover in the Hubbard model\cite{Micnus}. In a uniform Fermi gas with no lattice potential, $T_{\rm c}$ approaches the constant value $T_{\rm c}=0.218\varepsilon_{\rm F}$ in the BEC limit\cite{Melo,Haussmann94,Ohashi}. 
\par
We note that the decrease of $T_{\rm c}$ in the BEC regime does not mean the small binding energy of a bound molecule in this regime. As shown in Fig.\ref{fig4}, while the chemical potential $\mu$ is almost equal to the Fermi energy $\varepsilon_{\rm F}$ at $U=0$ (apart from weak temperature effect), it becomes smaller than the bottom of the band ($=-6t$) in the strong coupling regime. This means the existence of a {\it finite energy gap} $E_{\rm g}=|\mu|-6t$ in the Fermi single-particle excitations in the BEC regime. Since $E_{\rm g}$ is directly related to the dissociation energy of a bound molecule, we find that the binding energy $E_{\rm b}$ continues to increase even in the BEC regime (although $T_{\rm c}$ decreases as shown in Fig.\ref{fig3}).  
\par
To understand physics behind the decreasing $T_{\rm c}$ in the BEC regime, it is helpful to derive an effective model valid for this regime. To do this, we note that in the BEC regime, tightly bound molecules have been already formed above $T_{\rm c}$. In this case, as pointed out in Ref. \cite{Nozieres}, molecular motion is accompanied by virtual dissociation, because each atoms in a molecule has to move one by one in the Hubbard model. In addition, this virtual dissociation also leads to a repulsive interaction between molecules\cite{Nozieres}. Including these effects within the second order perturbation in terms of the hopping $t$, we obtain the effective Hamiltonian\cite{robasz81},
\begin{eqnarray}
H_{\rm eff}=-{2t^2 \over U}\sum_{(i,j)}[b^\dagger_i b_j+{\rm h.c.}]
+{4t^2 \over U}\sum_{(i,j)}n^{\rm B}_i n^{\rm B}_j
-\mu_{\rm B}\sum_i n^{\rm B}_i,
\label{eq19}
\end{eqnarray}
where $b^\dagger_i=c^\dagger_{i\uparrow}c^\dagger_{i\downarrow}$
describes a molecule at the $i$-th lattice site, and $n^{\rm
B}_i=(n_{i\uparrow}+n_{i\downarrow})/2$ gives the number of molecules
under the assumption that all the atoms form small on-site bound
pairs. $\mu_{\rm B}=2\mu+U+2zt^2/U$ is the molecular chemical potential to control the molecular density. In Eq. (\ref{eq19}), double occupancy of molecules is forbidden due to the Pauli's exclusion principle of Fermi atoms in them. Noting this and commutation relations, $[b_i^\dagger,b_i]=2(n_i^{\rm B}-1/2)$, $[n_i^{\rm B}-1/2,b_i^{\dagger}]=b_i^{\dagger}$, and $[n_i^{\rm B}-1/2,b_i]=-b_i$, we can map Eq. (\ref{eq15}) onto the $S=1/2$ Heisenberg model, by replacing $(b_i^\dagger,b_i,n_i^{\rm B}-1/2)$ with $((-1)^iS_i^+,(-1)^iS_i^-,S_i^z)$,
\begin{equation}
H_{\rm eff}
=J\sum_{(i,j)}{\bf S}_i\cdot{\bf S}_j-h_{\rm B}\sum_iS_i^z.
\label{eq20}
\end{equation}
Here, $J=4t^2/U$ is an exchange interaction and $h_{\rm B}=\mu_{\rm B}-2zt^2/U$ works as an external magnetic field. Since $S_i^z=-1/2$ and $S_i^z=+1/2$ state in Eq. (\ref{eq16}), respectively, correspond to a vacant and occupied site in the effective model in Eq. (\ref{eq19}), the half-filling case is described by setting $h_{\rm B}=0$ in Eq. (\ref{eq20}). Equation (\ref{eq20}) clearly shows that the `N\'{e}el temperature $T_{\rm N}$' (which corresponds to $T_{\rm c}$ in the original Hubbard model) is lower for larger $U$, consistent with the decreasing $T_{\rm c}$ in the strong-coupling BEC regime shown in Fig.\ref{fig3}.
\par
The N\'{e}el temperature of the Heisenberg model has been studied by various methods. It has been shown that the mean-field result $T_{\rm N}=6t^2/U$ is suppressed by spin fluctuations to be $T_{\rm N}=3.59t^2/U$ (high temperature expansion)\cite{rush63} and $T_{\rm N}=3.78t^2/U$ (quantum Monte Carlo method)\cite{sand98}. In the present calculation, we obtain $T_{\rm c}\simeq 1.5t^2/U < (3.59\sim3.78)t^2/U$ in the BEC regime, showing that the present calculation based on SCTA underestimates $T_{\rm c}$ in the BEC regime. 
\par
One reason for this discrepancy is the approximate treatment of the interaction between molecules. To see this, we consider the BCS-BEC crossover problem within the Gaussian fluctuation theory developed by Nozi\`eres and Schmitt-Rink\cite{Nozieres}. (We refer this theory as the NSR theory in the following.) In the NSR theory, one also solves the coupled equations (\ref{eq10}) and (\ref{eq11}), where the single-particle Green's function $G$ and the correlation function $\Pi_{\rm pp}$ are now replaced by
\begin{eqnarray}
G({\bm p},i\omega_m)=G_0(\bm{k},i\omega_m)
+G_0(\bm{k},i\omega_m) \Sigma_{\rm pp}(\bm{k},i\omega_m)
 G_0(\bm{k},i\omega_m),
\label{eq21}
\end{eqnarray}
\begin{eqnarray}
\Pi_{\rm pp}({\bm q},i\nu_n)={1 \over \beta}
\sum_{\bm{k},i\omega_m}
G_0(\bm{k},i\omega_m)G_0(\bm{q}-\bm{k},i\nu_n-i\omega_m),
\label{eq22}
\end{eqnarray}
where $G_0({\bf k},i\omega_m)^{-1}=i\omega_m-\varepsilon_{\bf k}+\mu$ is the Green's function in a free Fermi gas. The self-energy part in Eq. (\ref{eq21}) is given by
\begin{eqnarray}
\Sigma_{\rm pp}(\bm{k},i\omega_m)
={1 \over \beta}\sum_{\bm{q},i\nu_n}\Gamma_{\rm pp}(\bm{q},i\nu_n)
G_0(\bm{q}-\bm{k},i\nu_n-i\omega_m).
\label{eq23}
\end{eqnarray}
The particle-particle scattering vertex $\Gamma_{\rm pp}$ is given by Eq. (\ref{eq4}), where $\Pi_{\rm pp}$ is replaced by Eq.(\ref{eq22}). 
\par

%%%%%%%%%%%%%%%%%%%%%%%%%%%%%%%%%%%%%%%%%%%%%%%%%%%%%%%%%%%%%%%%%%%%%%%%%%%%%%%
\begin{figure}
\includegraphics[scale=0.7]{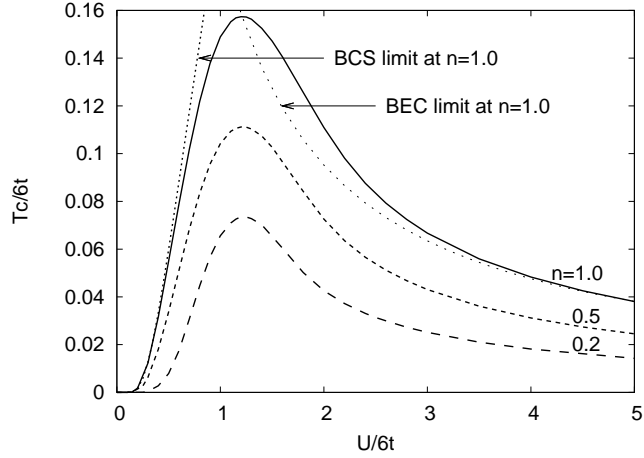}
\caption{Calculated $T_{\rm c}$ within the NSR theory. `BCS limit' is the mean-filed BCS result with $\mu=\varepsilon_{\rm F}$. `BEC limit' shows $T_{\rm c}$ of an ideal molecular Bose gas. 
\label{fig5}
}
\end{figure}
%%%%%%%%%%%%%%%%%%%%%%%%%%%%%%%%%%%%%%%%%%%%%%%%%%%%%%%%%%%%%%%%%%%%%%%%%%%%%%%
\par
Figure \ref{fig5} shows the calculated $T_{\rm c}$ based on the NSR
theory. Comparing this result with Fig.\ref{fig3}, we find that the NSR
theory gives higher $T_{\rm c}$. Since the NSR theory is a low density
approximation\cite{Nozieres} (Note that only the free propagator $G_0$ is used and the self-energy correction is only taken into account to
the first order in the number equation (\ref{eq10}).), the difference
between the two is more remarkable in higher filling cases.
\par
In the BEC limit of the NSR theory, Eq. (\ref{eq10}) gives $\mu=-U/2$, and the number equation (\ref{eq11}) reduces to the condition for BEC in an ideal Bose gas, 
\begin{equation}
{n \over 2}=\sum_{\bf k}
{1 \over e^{\beta(E_{\bf q}^{\rm B}-{\bar \mu}_{\rm B})}-1},
\label{eq24}
\end{equation}
where $E_{\bf q}^{\rm B}=-2(2t^2/U)[\cos q_x+\cos q_y+\cos q_z]$ and ${\bar \mu}_{\rm B}=-6(2t^2/U)$. Indeed, Fig.\ref{fig5} shows that $T_{\rm c}$ obtained from Eq. (\ref{eq24}) well describes the NSR result in the strong-coupling regime. Noting that the kinetic energy $E_{\bf q}^{\rm B}$ can be also obtained from the first term in Eq. (\ref{eq19}) when one regards $b_i$ as a boson operator, we find that the NSR theory ignores the repulsive interaction between molecules given by the second term in Eq. (\ref{eq19}). 
\par
In a uniform Fermi gas with no lattice potential, Haussmann pointed out that, in the BEC regime, SCTA includes the interaction between molecules within the Born approximation\cite{Haussmann94}. This molecular interaction can be written as $U_{\rm B}=4\pi a_{\rm B}/M_{\rm B}$, where $M_{\rm B}$ is a molecular mass. The $s$-wave molecular scattering length $a_{\rm B}$ is related to the $s$-wave atomic scattering length $a_s$ as $a_B=2a_s$. Even in the presence of the lattice, the molecular interaction is expected to be included within the same approximation level. Thus, we find that the lower $T_{\rm c}$ within SCTA than the NSR result originates from the molecular interaction. Namely, the molecular interaction lowers $T_{\rm c}$ in the BEC regime in the lattice system. 
\par
Recent work\cite{Strinati,Petrov,Gio,Ohashi3} on a uniform Fermi gas has clarified that, when one carefully treats higher order molecular scattering processes and a finite value of molecular binding energy, $a_{\rm B}$ reduces to $a_{\rm B}=0.6a_s<2a_s$. This clearly indicates {\it overestimate} of the magnitude of molecular interaction in SCTA. When we apply this discussion to the present lattice system, one reason for the underestimate of $T_{\rm c}$ $( =1.5t^2/U < (3.59\sim3.78)t^2/U)$ is expected to be the overestimate of the molecular interaction. When one could correct this point, $T_{\rm c}$ would be higher to be close to the `N\'{e}el temperature' of the Heisenberg model in Eq. (\ref{eq20}). This improvement is an interesting problem; however, in this paper, leaving this as a future problem, we treat pairing fluctuations within SCTA and discuss effects of CDW and SDW fluctuations in the next section.
\par

%%%%%%%%%%%%%%%%%%%%%%%%%%%%%%%%%%%%%%%%%%%%%%%%%%%%%%%%%%%%%%%%%%%%%%%%%%%%%%%
\begin{figure}
\includegraphics[scale=0.7]{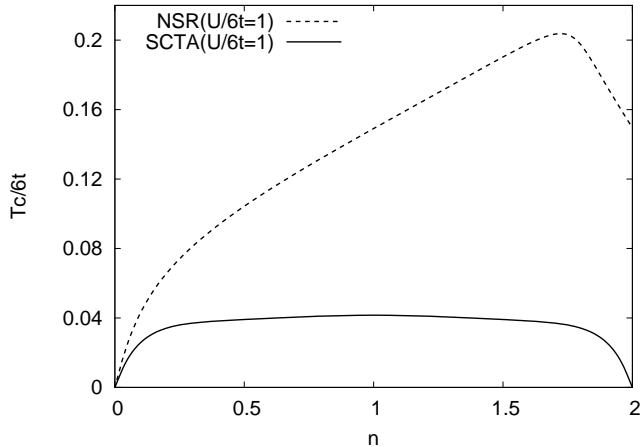}
\caption{$T_{\rm c}$ as a function of the filling number $n$. The result within the SCTA and that within the NSR theory are compared. We set $U/6t=1$.
\label{fig6}
}
\end{figure}
%%%%%%%%%%%%%%%%%%%%%%%%%%%%%%%%%%%%%%%%%%%%%%%%%%%%%%%%%%%%%%%%%%%%%%%%%%%%%%%
\begin{figure}
\includegraphics[scale=0.7]{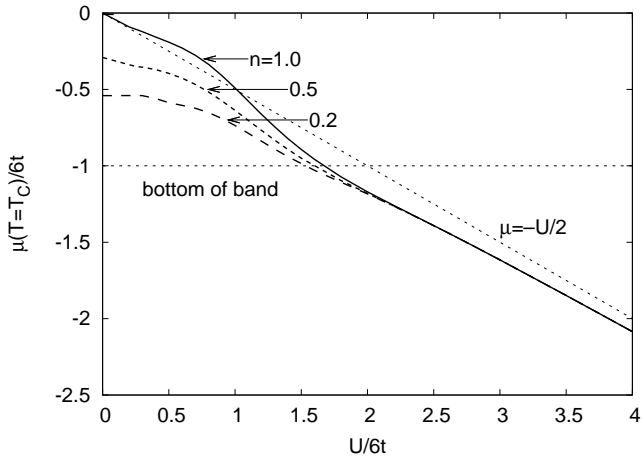}
\caption{Chemical potential $\mu$ at $T_{\rm c}$ calculated within the NSR theory. 
\label{fig7}
}
\end{figure}
%%%%%%%%%%%%%%%%%%%%%%%%%%%%%%%%%%%%%%%%%%%%%%%%%%%%%%%%%%%%%%%%%%%%%%%%%%%%%%%
\par
Before ending this section, we briefly note that the NSR theory does not
satisfy the particle-hole symmetry, when it is applied to the Hubbard
model. As shown in Fig.\ref{fig6}, the calculated $T_{\rm c}$
based on the NSR theory is unphysical around $n=2$. (Note that the system
must be a band insulator at $n=2$, leading to vanishing $T_{\rm c}$.). We also find that the required symmetric filling dependence of $T_{\rm c}$ with respect to $n=1$ is not obtained within the NSR theory.
In addition, although the chemical potential must satisfy $\mu=-U/2$ at $n=1$ due to the particle-hole symmetry\cite{Valkov}, the NSR result satisfies it only in the BEC limit $U\to\infty$, as shown in Fig.\ref{fig7}. In contrast, in addition to the symmetric filling dependence of $T_{\rm c}$, SCTA can also reproduce the exact result $\mu=-U/2$ at $n=1$ over the entire BCS-BEC crossover. (See Fig.\ref{fig4}.) We emphasize that satisfying these required conditions is important in any consistent theory.
\par
%%%%%%%%%%%%%%%%%%%%%%%%%%%%%%%%%%%%%%%%%%%%%%%%%%%%%%%%%%%%%%%%%%%%%%%%%%%%%%%
\begin{figure}
\includegraphics[scale=0.7]{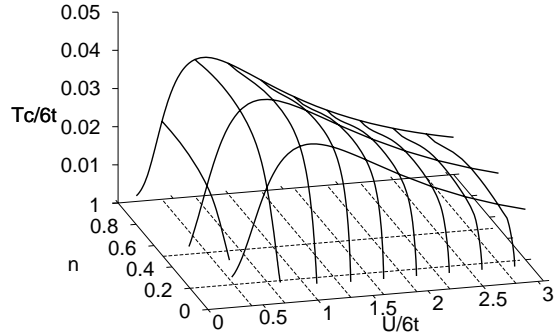}
\caption{Superfluid phase transition temperature $T_{\rm c}$ as a
 function of pairing interaction $U$ and filling number $n$. In this
 figure, and in Fig.\ref{fig9}, we include CDW and SDW fluctuations
 described by $\Sigma_{\rm ph}^{d,s}$ in addition to pairing
 fluctuations.
In comparison with Fig.\ref{fig4}, $T_{\rm c}$ around $n=1$ are suppressed in the strong coupling regime.
\label{fig8}
}
\end{figure}
%%%%%%%%%%%%%%%%%%%%%%%%%%%%%%%%%%%%%%%%%%%%%%%%%%%%%%%%%%%%%%%%%%%%%%%%%%%%%%%
\begin{figure}
\includegraphics[scale=0.7]{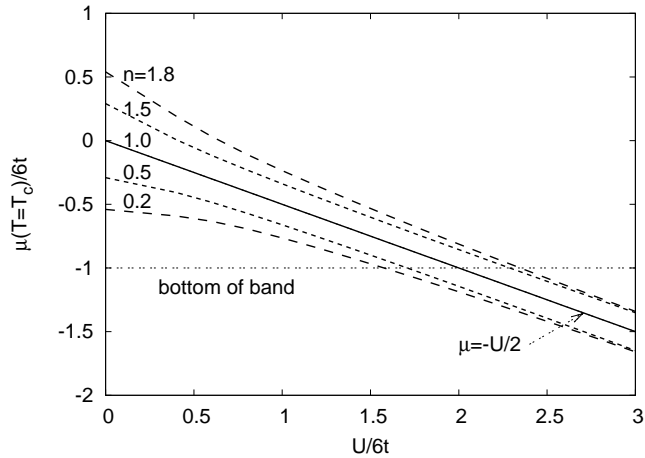}
\caption{Chemical potential $\mu$ as a function of $U$ at $T_{\rm
 c}$ when we consider CDW and SDW fluctuations in addition to pairing fluctuations. Self-consistent
 treatment of CDW and SDW fluctuations does not break the particle-hole symmetry
 condition $\mu=-U/2$ at $n=1$.
\label{fig9}
}
\end{figure}
%%%%%%%%%%%%%%%%%%%%%%%%%%%%%%%%%%%%%%%%%%%%%%%%%%%%%%%%%%%%%%%%%%%%%%%%%%%%%%%

%%%%%%%%%%%%%%%%%%%%%%%%%%%%%%%%%%%%%%%%%%%%%%%%%%%%%%%%%%%%%%%%%%%%%%%%%%%%%%%
\begin{figure}
\includegraphics[scale=0.7]{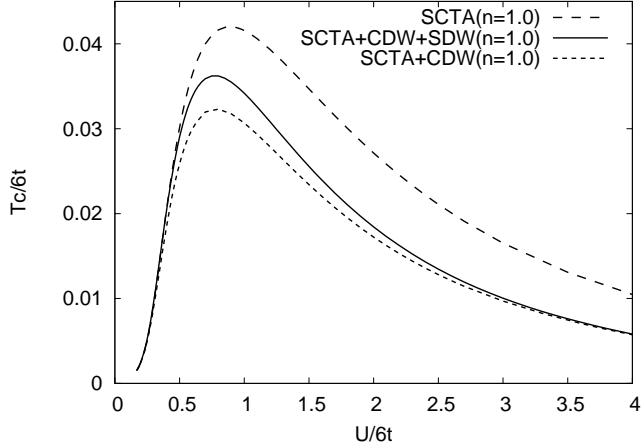}
\caption{Effects of CDW and SDW fluctuations on $T_{\rm c}$ at the half-filling $n=1$. SCTA: Pairing fluctuations are only included. SCTA+CDW+SDW: Pairing, CDW, and SDW fluctuations are all included. SCTA+CDW: Pairing and CDW fluctuations are taken into account. These abbreviations are also used in Figs.\ref{fig11} and \ref{fig12}. 
\label{fig10}
}
\end{figure}
%%%%%%%%%%%%%%%%%%%%%%%%%%%%%%%%%%%%%%%%%%%%%%%%%%%%%%%%%%%%%%%%%%%%%%%%%%%%%%%

%%%%%%%%%%%%%%%%%%%%%%%%%%%%%%%%%%%%%%%%%%%%%%%%%%%%%%%%%%%%%%%%%%%%%%%%%%%%%%
\begin{figure}
\includegraphics[scale=0.7]{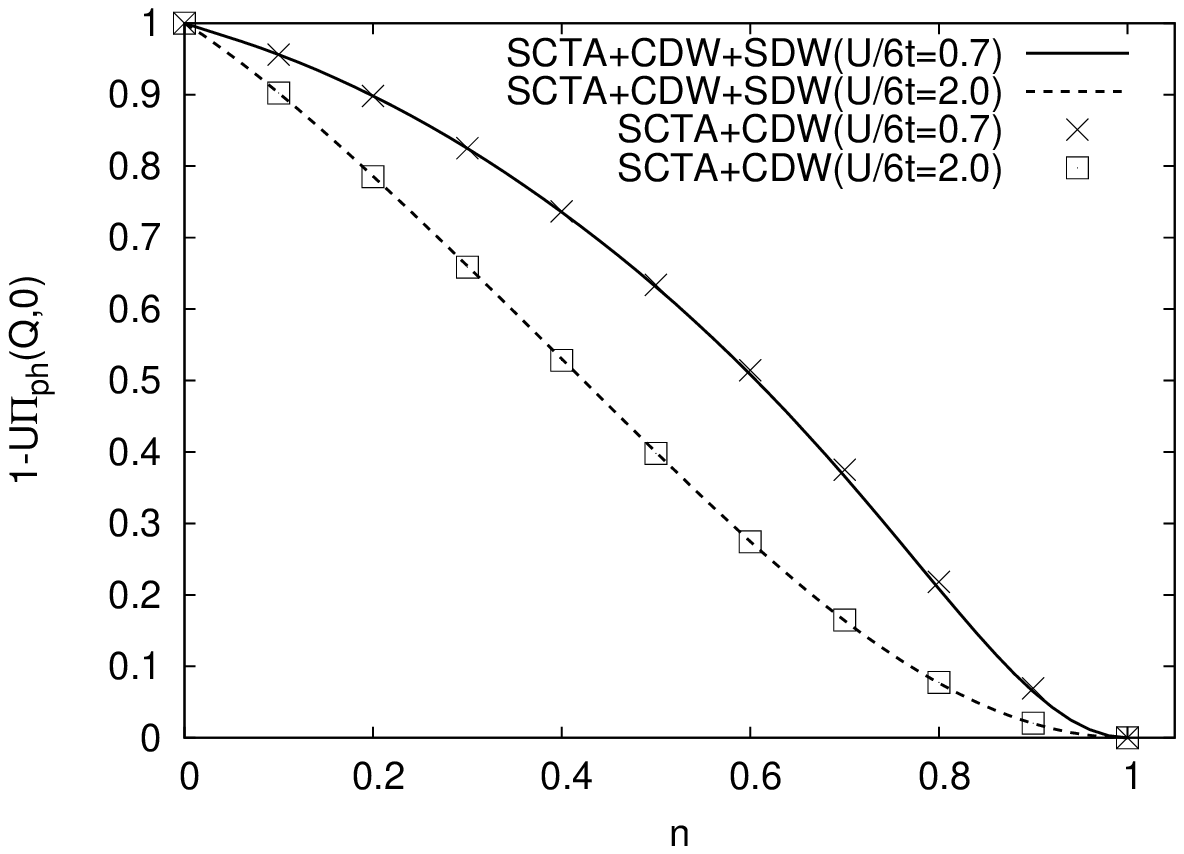}
\caption{Denominator of the static charge susceptibility $\chi({\bf Q})$ in Eq. (\ref{eq12}) at the superfluid phase transition temperature $T_{\rm c}$. The CDW instability is determined when $1-U\Pi_{\rm ph}(\bm{q}=\bm{Q},\omega =0)=0$ is realized.
\label{fig11}
}
\end{figure}
%%%%%%%%%%%%%%%%%%%%%%%%%%%%%%%%%%%%%%%%%%%%%%%%%%%%%%%%%%%%%%%%%%%%%%%%%%%%%%

\section{Effects of CDW and SDW fluctuations}

Figures \ref{fig8} and \ref{fig9} show calculated $T_{\rm c}$ and $\mu(T_{\rm c})$ respectively in the BCS-BEC crossover, when the CDW ($\Sigma_{\rm ph}^{\rm d}$) and SDW fluctuations ($\Sigma_{\rm ph}^{\rm s}$) are both taken into account. As shown in Fig. \ref{fig9}, the required condition $\mu=-U/2$ at $n=1$ is still satisfied when one includes CDW and SDW fluctuations within FLEX.
\par
Comparing Fig.\ref{fig4} with Fig.\ref{fig9}, one finds that effects of CDW and SDW fluctuations on the chemical potential $\mu$ are weak. In contrast, from the comparison of Fig.\ref{fig3} and Fig.\ref{fig8}, $T_{\rm c}$ is found to be suppressed near the half-filling when these fluctuations are taken into account. 
\par
To see the suppression of $T_{\rm c}$ more clearly, we show $T_{\rm c}$ at $n=1$ in Fig.\ref{fig10}. At the half-filling, the CDW instability occurs simultaneously (which is confirmed by vanishing denominator of the charge susceptibility $\chi({\bf Q})$ as shown in Fig.\ref{fig11}), which leads to the remarkable suppression of $T_{\rm c}$. However, in contrast to the two-dimensional case, where $T_{\rm c}$ vanishes at $n=1$\cite{Scalettar}, we still obtain a finite $T_{\rm c}$ even at the half-filling. 
\par
Although SDW fluctuations are weak in the attractive Hubbard model, we still find their effects around $U/6t\sim 1$ in Fig.\ref{fig10}. From the comparison of the result referred to as 'SCTA+CDW+SDW' with 'SCTA+CDW' in Fig.\ref{fig10}, we find that SDW fluctuations weaken the suppression of $T_{\rm c}$ by CDW fluctuations. 
\par
In this intermediate coupling regime, since the binding energy of a Cooper pair is not very strong, (pseudo)spin degrees of freedom still remains, which contribute to SDW fluctuations. As one approaches the strong-coupling regime, these spin degrees from freedom disappear due to the formation of singlet pairs. Indeed, in Fig.\ref{fig10}, the two results, 'SCTA+CDW+SDW' and 'SCTA+CDW', give almost the same $T_{\rm c}$ when $U/6t\gesim 3$.
\par
%%%%%%%%%%%%%%%%%%%%%%%%%%%%%%%%%%%%%%%%%%%%%%%%%%%%%%%%%%%%%%%%%%%%%%%%%%%%%%
\begin{figure}
\includegraphics[scale=0.7]{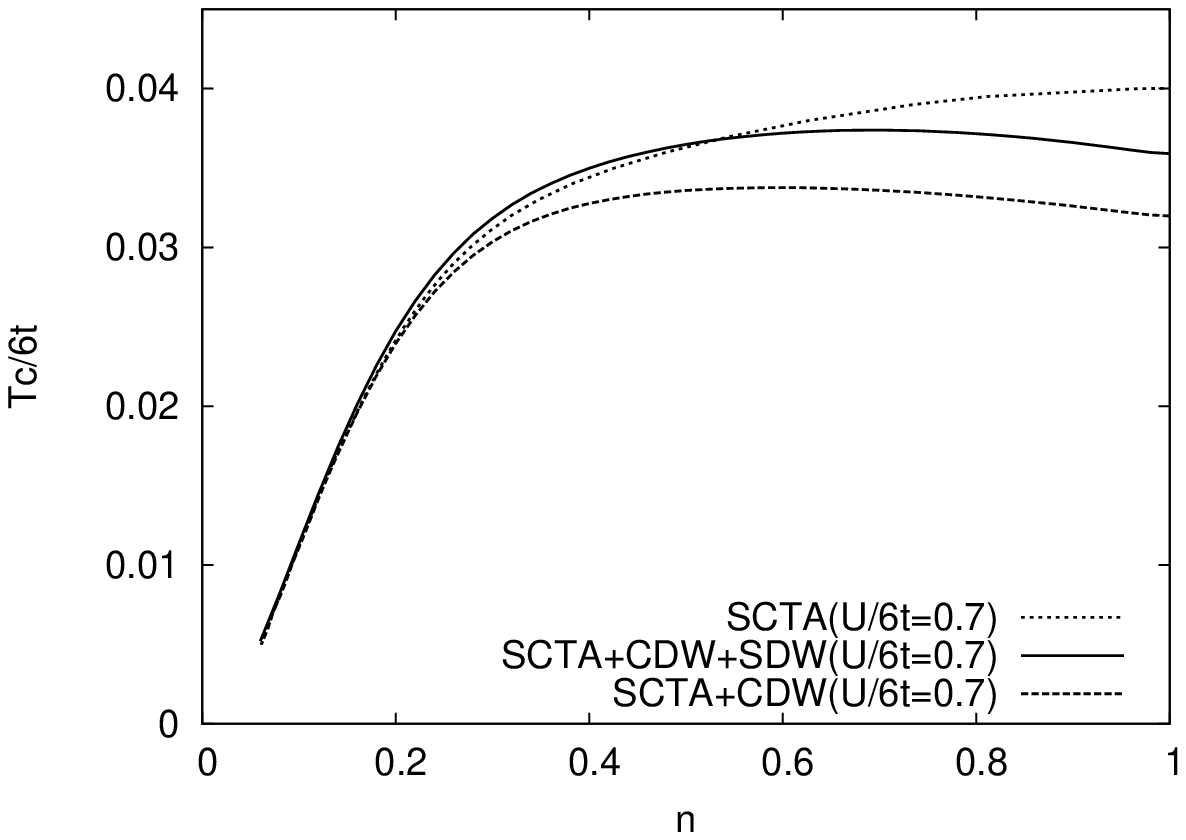}\\
\includegraphics[scale=0.7]{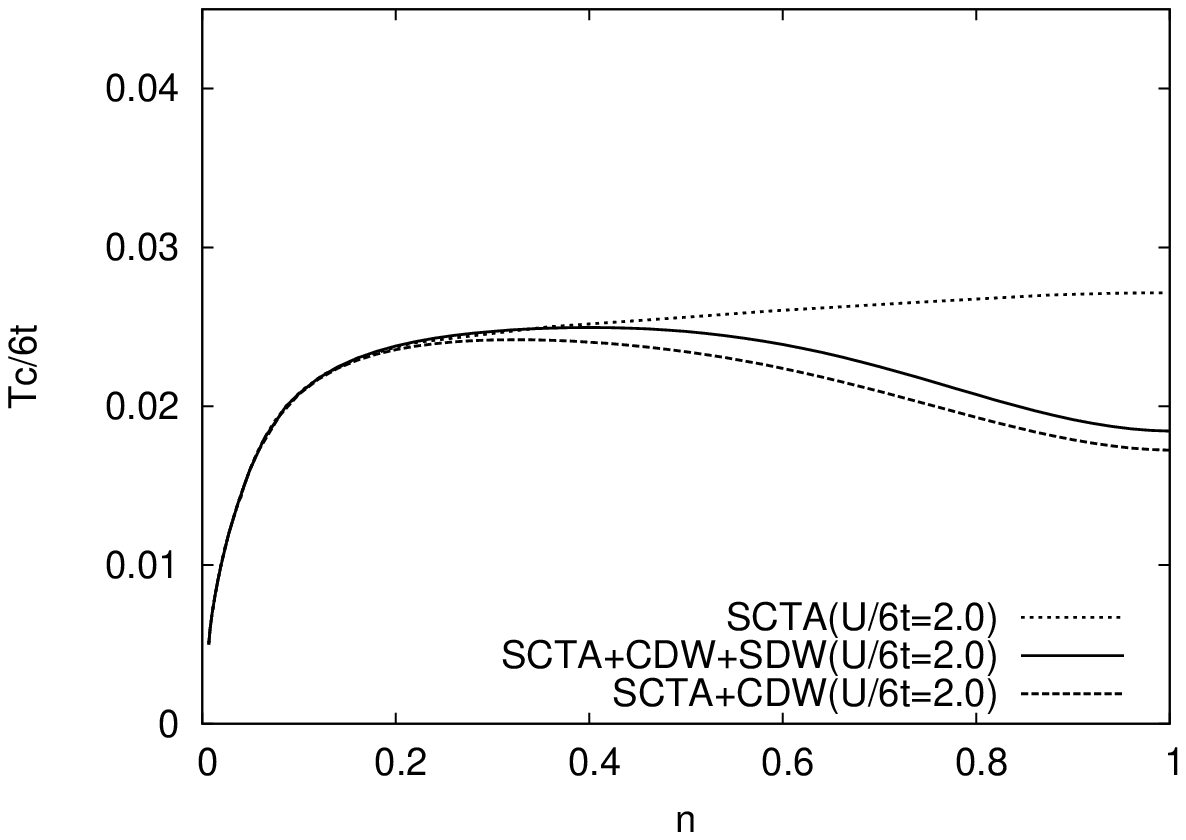}\\
\caption{Filling dependence of the superfluid transition temperature $T_{\rm c}$. Upper and lower panels show the cases of the weak-coupling BCS regime ($U/6t=0.7$) and strong-coupling BEC regime ($U/6t=2$), respectively.
\label{fig12}
}
\end{figure}
%%%%%%%%%%%%%%%%%%%%%%%%%%%%%%%%%%%%%%%%%%%%%%%%%%%%%%%%%%%%%%%%%%%%%%%%%%%%%%
\par
Although the degeneracy of the superfluid state and CDW is absent when
$n\ne 1$, we can still expect strong influence of CDW near the
half-filling due to strong enhancement of charge susceptibility shown in
Fig.\ref{fig11}. Indeed, as shown in Fig.\ref{fig12}, CDW fluctuations
suppress $T_{\rm c}$ near $n=1$. (Compare `SCTA' with `SCTA+CDW+SDW' in
Fig.\ref{fig12}.) The maximum $T_{\rm c}$ is thus obtained, not at the
half-filling, but away from the half-filling. Although SDW fluctuations
enhance $T_{\rm c}$, the overall behavior is unchanged. (Compare
`SCTA+CDW+SDW' with `SCTA+CDW' in Fig.\ref{fig12}.) Since spin degrees
of freedom are almost absent in the strong-coupling regime, effects of
SDW fluctuations are weaker in the lower panel (BEC regime) than the
upper panel (BCS$\sim$crossover regime) in Fig.\ref{fig12}.

\par
%%%%%%%%%%%%%%%%%%%%%%%%%%%%%%%%%%%%%%%%%%%%%%%%%%%%%%%%%%%%%%%%%%%%%%%%%%%%%%
\par
\section{Summary}
To summarize, we have discussed the superfluid phase transition in the
BCS-BEC crossover regime of a two-component Fermi gas loaded on a
three-dimensional optical lattice. Treating this system as the
attractive Hubbard model, we calculated the superfluid phase transition
temperature $T_{\rm c}$, including pairing fluctuations within the
self-consistent $t$-matrix theory, as well as CDW and SDW fluctuations
within the fluctuation exchange approximation. We determined $T_{\rm c}$
and the Fermi chemical potential $\mu$ self-consistently as functions of
the pairing interaction $U$ and filling number $n$ in the BCS-BEC
crossover region, by solving the equation for $T_{\rm c}$, together with
the number equation.
\par
CDW fluctuations are strong near the half-filling due to the nesting
property of the Fermi surface at $n=1$. These strong fluctuations
remarkably decrease the superfluid phase transition temperature $T_{\rm
c}$ around $n=1$. As a result, the maximum $T_{\rm c}$ is obtained, not
at the half-filling, but away from the half-filling. We also showed
that, although SDW fluctuations are weak in the attractive Hubbard
model, they still affect $T_{\rm c}$ to increase slightly in the intermediate coupling region
$U/6t\sim 1$. 
\par
We have also discussed the validity of SCTA. 
Our theory satisfies the required condition associated with the particle-hole symmetry of the
Hubbard model which is not derived from NSR theory.
Therefore, it would be a good starting point to improve the BCS-BEC crossover theory in optical lattices.
On the other hand, we showed that this
approximation underestimates $T_{\rm c}$ in the BEC regime. As a key to
understand this, we pointed out the importance of a repulsive
interaction between molecules. Since the Gaussian fluctuation theory
(which completely ignores the molecular interaction at $T_{\rm c}$)
largely {\it overestimates} $T_{\rm c}$, we expect that SCTA
overestimates effects of the molecular interaction. Indeed, the
overestimate of the molecule interaction within SCTA has been pointed
out\cite{Haussmann94,Strinati,Petrov,Gio,Ohashi3} in a uniform Fermi gas
with no lattice potential. Inclusion of the correct value of the
molecular interaction in the present theory is our future
problem. 
\par
So far, the superfluid Fermi gas in an optical lattice has been realized
when the lattice potential is weak\cite{chin06}. To really realize the
Hubbard model in a cold gas of Fermi atoms, stronger optical lattice
potentials must be used, which, however, inevitably leads to a small
nearest-neighbor hopping $t$, as well as low $T_{\rm c}$. Thus, to
realize Fermi superfluids in such a difficult situation, our results
indicate that the filling number should be set to be away from $n=1$ (to avoid the suppression of $T_{\rm
c}$ by CDW fluctuations) so that one can reach $T_{\rm c}$ as easy as
possible under a given experimental condition. Since the Hubbard model
is a fundamental model in condensed matter physics, realization of Fermi
superfluid in the Hubbard model produced by strong optical lattice
potential would be great challenge in cold atom physics.
%%%%%%%%%%%%%%%%%%%%%%%%%%%%%%%%%%%%%%%%%%%%%%%%%%%%%%%%%%%%%%%%%%%%%%%%%%%%%%%
\acknowledgments
This work was supported by a Grant-in-Aid for Scientific
Research of Priority Area ``Physics of New Quantum Phase in Superclean
Materials'' from the Ministry of Education, Culture, Sports, Science and
Technology of Japan (MEXT). The authors were also supported by MEXT
(No. 19340099 (H. T. and K. M.), and 19540420 (Y. O.)).
%%%%%%%%%%%%%%%%%%%%%%%%%%%%%%%%%%%%%%%%%%%%%%%%%%%%%%%%%%%%%%%%%%%%%%%%%%%%%%%
\appendix
\section{Application of FFT to Frequency summations}
\par
In this Appendix, we explain how to use FFT algorithm in transforming
between the Matsubara frequency and the imaginary time. As an example, we consider the single-particle Green's function $G$ here. However, the method explained in this appendix can be also used in calculating the self-energies as well as correlation functions.
\par
We introduce a Matsubara frequency cutoff $\omega_{\rm max}=\pi T(2n_{\rm max}+1)$ and
evaluate the Green's function by using the Fourier transformation.
By introducing this cutoff frequency, the Fourier transformation from the Matsubara frequency into imaginary time can be rewritten as (suppressing the variables ${\bf k}$ and ${\bf r}$),
\begin{eqnarray}
G(\tau)&=&T\sum_{m=-n_{\rm max}-1}^{n_{\rm max}}G(i\omega_m)e^{-i\omega_m\tau}\nonumber\\
&=&2T\sum_{m=0}^{n_{\rm max}}\bigl({\rm Re}[G(i\omega_m)]\cos{\omega_m\tau}+{\rm Im}[G(i\omega_m)]\sin{\omega_m\tau}\bigr),\label{A1}
\end{eqnarray}
where we have used the analytic property $G(-i\omega_m)=G^{*}(i\omega_m)$, as well as the fact that $G(\tau)$ is a real function. Replacing $\tau$ by $\beta-\tau$ in Eq. (\ref{A1}), we have
\begin{eqnarray}
G(\beta-\tau)
&=&2T\sum_{m=0}^{n_{\rm max}}\bigl(-{\rm Re}[G(i\omega_m)]\cos{\omega_m\tau}+{\rm Im}[G(i\omega_m)]\sin{\omega_m\tau}\bigr).\label{A2}
\end{eqnarray}
Thus, one may only consider the region $0\le\tau<\beta/2$ in executing the cosine (sine) Fourier transformation from the Matsubara frequency into imaginary time. In numerical calculations, we divide the region $0\le\tau<\beta/2$ into $n_{\rm max}$ cells and use FFT method.
\par
When we calculate the inverse Fourier transformation, we meet the problem that the expected high-frequency behavior $G({\bm k},i\omega_m)\sim 1/i\omega_m$ is not obtained because of the introduced cutoff frequency $\omega_{\rm max}$. To avoid this problem,
we rewrite the inverse Fourier transformation in the form,
\begin{eqnarray}
G(i\omega_m)
=\int_0^\beta d\tau G(\tau)e^{i\omega_m\tau}
=\sum_{j=1}^{2n_{\rm max}}
\int_{\tau_{j-1}}^{\tau_j}d\tau G(\tau)e^{i\omega_m\tau}.
\label{ap1}
\end{eqnarray} 
Here, $\tau_j=\Delta\tau j$, where $\Delta\tau=\beta/2n_{\rm max}$. When we approximately write the Green's function in the region $\tau=[\tau_{j-1},\tau_j]$ as $G(\tau)\simeq G(\tau_{j-1})+(\tau-\tau_{j-1})\times(G(\tau_{j})-G(\tau_{j-1}))/\Delta\tau$, we can execute the integrals in Eq. (\ref{ap1}). The result is 
\begin{eqnarray}
G(i\omega_m)
&=&
{1 \over i\omega_m}[-G(\tau_{2n_{\rm max}}=\beta-\delta)-G(\tau_0=+\delta)]
+{1 \over \omega_m^2\Delta\tau}
\Bigl[
-\{G(\tau_1)-G(\tau_0)\}
\nonumber\\
&+&
\sum_{j=1}^{2n_{\rm max}-1}
\{2G(\tau_j)-G(\tau_{j+1})-G(\tau_{j-1})\}e^{i\omega_m\tau_j}
-\{G(\tau_{2n_{\rm max}})-G(\tau_{2n_{\rm max}-1})\}
\Bigr],\nonumber\\
\label{ap2}
\end{eqnarray} 
where $\delta$ is a infinitesimal positive number.
In Eq. (\ref{ap2}), because $G(\beta-\delta)=-G(-\delta)$, the first term remains finite due to the discontinuity of Green's function at $\tau=0$, giving the expected high frequency behavior ($\sim 1/i\omega_m$). We apply FFT to calculate the second term in (\ref{ap2}). We note that Eq. (\ref{ap2}) has been also derived in Ref.\cite{deisz02} by integration by parts.

%%%%%%%%%%%%%%%%%%%%%%%%%%%%%%%%%%%%%%%%%%%%%%%%%%%%%%%%%%%%%%%%%%%%%%%%%%%%%%%
% References

\end{document}